\documentclass[prl,a4paper,twocolumn,twoside,showpacs]{revtex4}
\usepackage{times}
\usepackage{graphicx,units}
\usepackage[usenames,dvipsnames]{color}
\usepackage{amssymb,stmaryrd}
\usepackage{pifont,wasysym}
\usepackage{natbib}
\newcommand{\Diamondblack}{\ding{117}}
\newcommand{\medbullet}{\ding{108}}

\usepackage[bookmarks=true,letterpaper=true,colorlinks=true,urlcolor=blue,linkcolor=blue,citecolor=blue]{hyperref}
\pacs{73.22.-f 71.15.Mb 73.21.-b}
\received{17 December 2009}
\begin{document}
\title{Trends in Metal Oxide Stability for Nanorods, Nanotubes, and Surfaces}
\author{D.~J.~Mowbray$^1$$^2$}
\author{J.~I.~Mart{\'{\i}}nez$^1$$^3$}
\author{F.~Calle-Vallejo$^1$} 
\author{J.~Rossmeisl$^1$}
\author{K.~S.~Thygesen$^1$}
\author{K.~W.~Jacobsen$^1$}
\author{J.~K.~N{\o}rskov$^1$}
\email{norskov@fysik.dtu.dk}
\affiliation{$^1$
Center for Atomic-scale Materials Design, Department of Physics, Technical University of Denmark, DK-2800 Kgs.~Lyngby, Denmark\\
$^2$Nano-Bio Spectroscopy group and ETSF Scientific Development Centre, Dpto.~F{\'{\i}}sica de Materiales, Universidad del Pa{\'{\i}}s Vasco, Centro de F{\'{\i}}sica de Materiales CSIC-UPV/EHU-MPC and DIPC,\\ Av.~Tolosa 72, E-20018 San Sebasti{\'{a}}n, Spain\\
$^3$Dpto.~de F{\'{\i}}sica Te{\'{o}}rica de la Materia Condensada, Universidad Aut{\'{o}}noma de Madrid, E-28049 Madrid, Spain
}
\begin{abstract}
The formation energies of nanostructures play an important role in determining their properties, including the catalytic activity. For the case of 15 different rutile and 8 different perovskite metal oxides, we find that the density functional theory (DFT) calculated formation energies of (2,2) nanorods, (3,3) nanotubes, and the (110) and (100) surfaces may be described semi-quantitatively by the fraction of metal--oxygen bonds broken and the bonding band centers in the bulk metal oxide.  
\end{abstract}

\maketitle

The search for cleaner and more sustainable forms of energy provides a strong impetus to the development of more affordable, active, selective and stable new catalysts to convert solar radiation into fuels \cite{DOE,Crabtree}.  Just as the Haber-Bosch process fueled the population explosion of the 20$^{\text{th}}$ century \cite{HaberBosch1,HaberBosch2}, it is now hoped that new catalytic processes will provide sustainable energy in the 21$^{\text{st}}$ century \cite{DOE,Crabtree}.

Metal oxides are used extensively as catalysts, electrocatalysts, and photo-electrocatalysts \cite{review,ElectrocatalysisRef24,ElectrocatalysisIrSnRuRef25,ElectrocatalysisIrTiRef26,ElectrocatalysisTiO2110Ref27,Rossmeisl200783,Rossmeisl2008,Fujishima1972,Khan2002,Hoffmann1995,Turner,FirstPhotocatalysis,Gratzel2001}.  One important property of oxides is their high stability in harsh oxidizing environments compared to their pure metal counterparts.

The structure of oxide nanoparticles may be determined by the surface energy.  For instance, for TiO$_2$ it has recently been shown that it is the surface energies which determine whether it takes the anatase or rutile structure at the nanoscale \cite{TiO2SurfaceEnergies,SurfaceEnthalpiesTiO2}. Typically, oxide catalysts are in the form of nanoparticles or highly porous materials. The catalytic properties of these materials are determined to a high degree by the surface \cite{Surfaces}, and control of the surface structure will allow control of the reactivity \cite{TiO2,TiO2Nanowires,TiO2AnataseNTs,TiO2Rev1}. However, formation energies of metal oxide surfaces are difficult to measure experimentally, and only a few values are available in the literature \cite{TiO2SurfaceEnergies,SurfaceEnthalpiesTiO2}.
 This makes the calculation of trends in surface and nanostructure energies an essential first step in understanding the properties of oxide catalysts, and the eventual design of novel catalytic materials.

\begin{figure}
\includegraphics[width=\columnwidth]{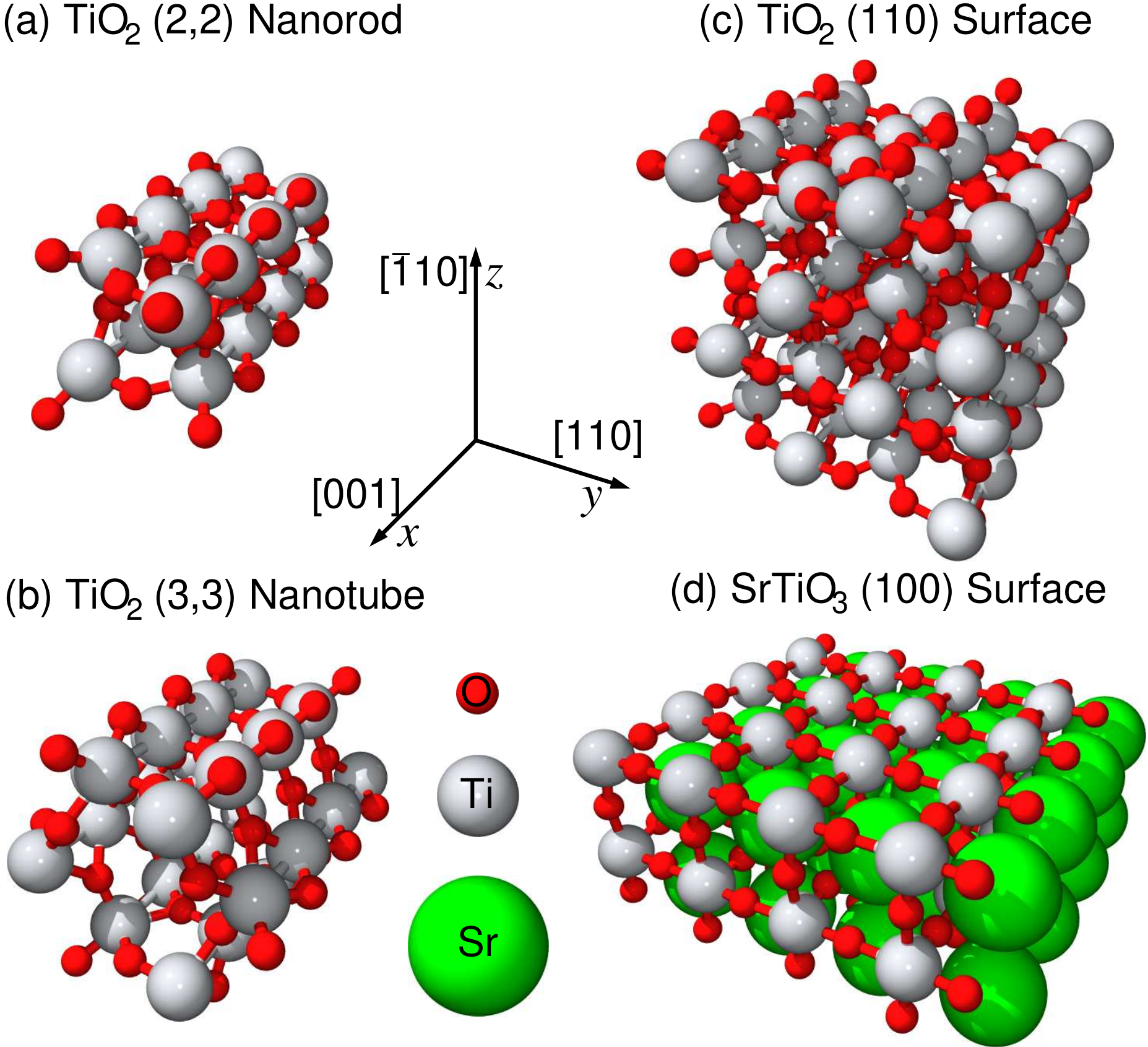}
\caption{Structural schematics for (a) TiO$_2$ (2,2) nanorod, (b) TiO$_2$ (3,3) nanotube, (c) TiO$_2$ (110) surface and (d) SrTiO$_3$ (100) TiO$_2$ and SrO terminated surfaces.  Axes and Miller indices for the rutile structures are shown for (a) and (c).}\label{Structures}
\end{figure}

\begin{figure*}
\includegraphics[width=\textwidth]{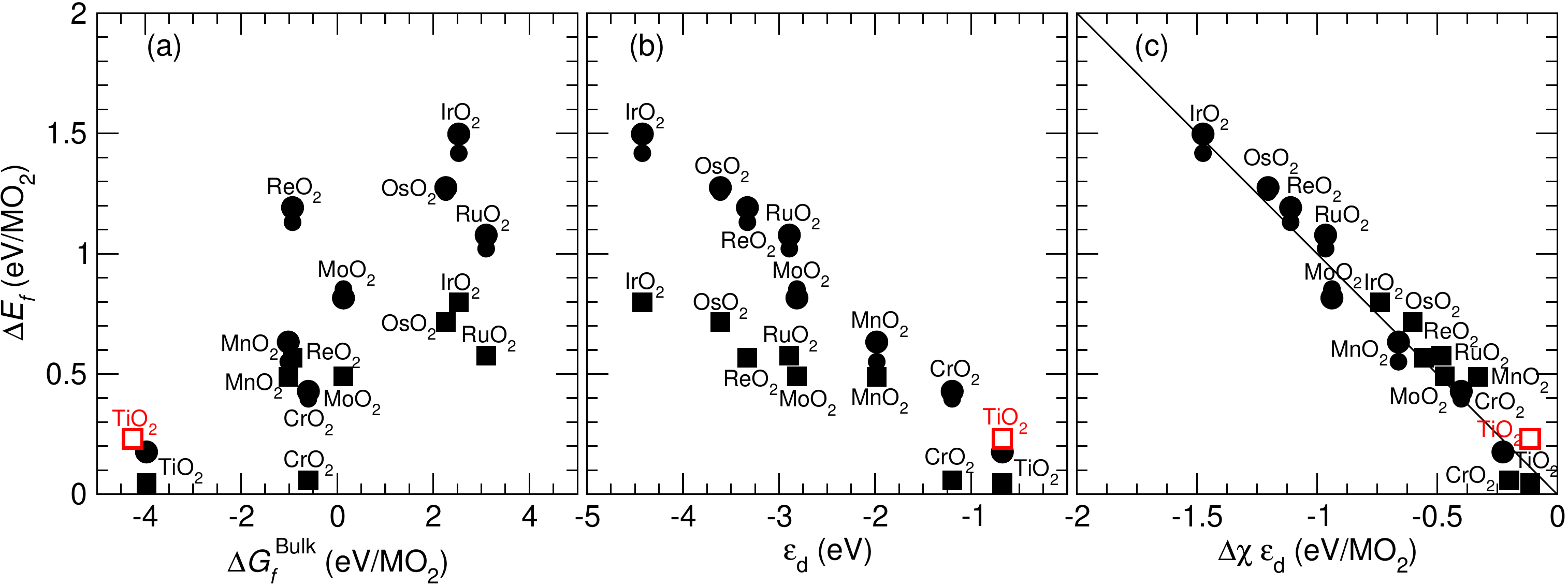}
\caption{Formation energy \(\Delta E_f\) of rutile metal oxide (2,2) nanorods ({\footnotesize\medbullet}), (3,3) nanotubes ({\medbullet}), and (110) surfaces ($\blacksquare$) in eV/MO$_{\text{2}}$ vs.~(a) bulk heat of formation \(\Delta G_f^{\text{Bulk}}\)for rutile metal oxides in eV/MO$_2$ 
from Ref.~\onlinecite{Martinez}, (b) $d$-band center \(\varepsilon_{d}\) for the bulk metal oxide relative to the Fermi energy in eV, and (c) fraction of M--O bonds broken times the $d$-band center \(\Delta \chi \varepsilon_{d}\) in eV/MO$_2$.  The experimental nanoparticle surface energy \cite{SurfaceEnthalpiesTiO2} and bulk formation energy \cite{CRCHandbook,Pourbaix} for rutile TiO$_2$ ({\color{red}{$\boxempty$}}) are provided for comparison.
}\label{Heats}
\end{figure*}

\begin{figure*}
\includegraphics[width=0.8\textwidth]{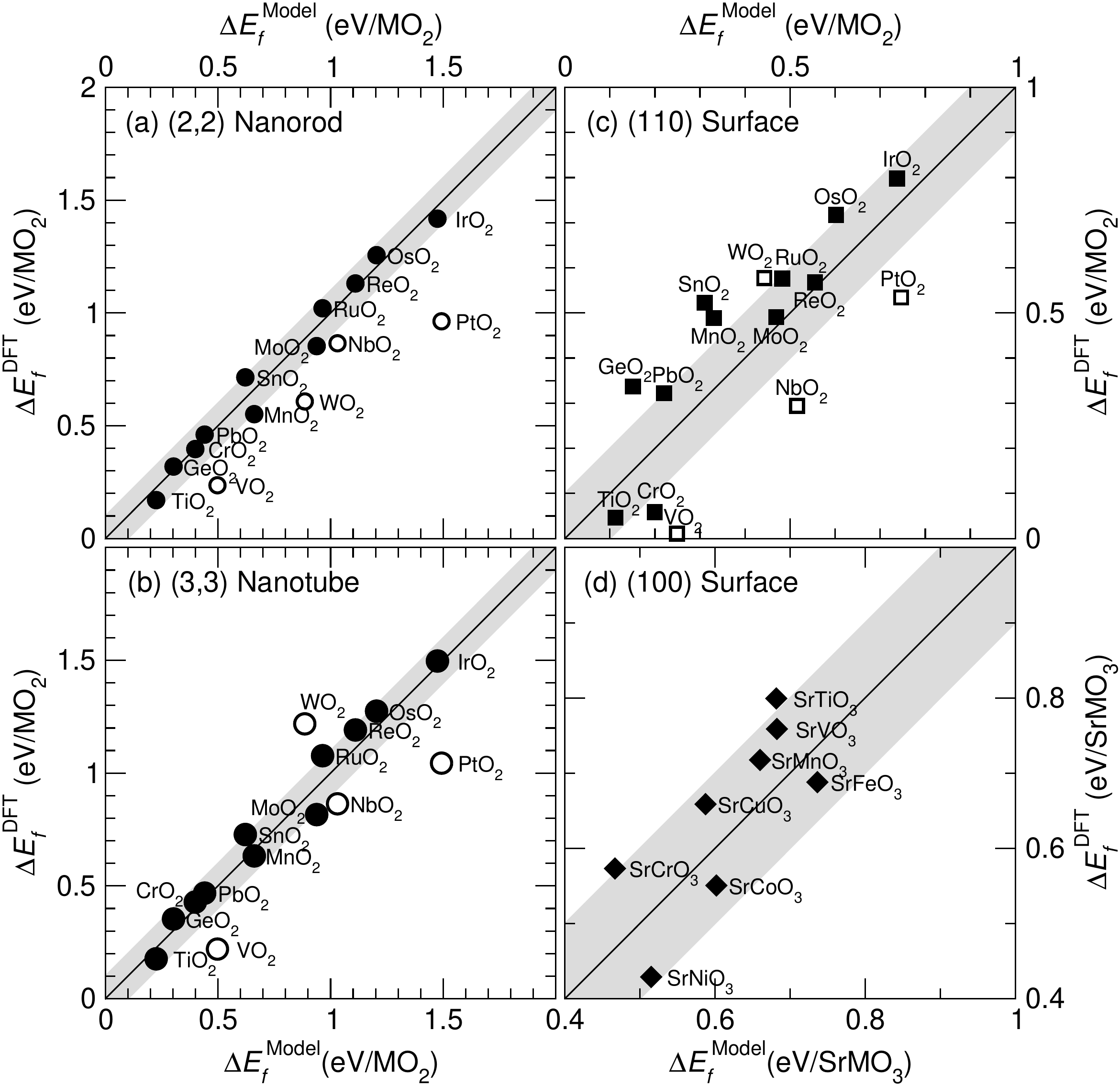}
\caption{DFT calculated formation energy \(\Delta E_{f}^{\text{DFT}}\) vs.~model prediction \(\Delta E_f^{\text{Model}}\) for (a) (2,2) nanorod ({\footnotesize\medbullet}), (b) (3,3) nanotube ({\medbullet}), and (c) (110) surface ($\blacksquare$) in eV/MO$_{\text{2}}$ for rutile and (d) (100) MO$_2$ and SrO terminated surfaces (\Diamondblack) in eV/SrMO$_3$ for perovskites. Open symbols denote metal oxides whose most stable phase is not rutile.  An error estimate of \(\pm\)0.1 eV (grey region) is provided for comparison.
}\label{DFT_vs_model}
\end{figure*}

In this study we provide DFT calculated (2,2) nanorod, (3,3) nanotube, and (110) surface formation energies for 15 different rutile metal oxides, along with the (100) surface formation energies for 8 different perovskite metal oxides \cite{TiO2}. Both the (2,2) nanorod and (3,3) nanotube structures are obtained by first rolling up a sheet of the material in the PtO$_2$ structure.  But due to its small diameter, the (2,2) nanorod has a bonding structure resembling that of the bulk metal after relaxation, as shown in Fig.~\ref{Structures}.  

All DFT calculations have been performed with the plane wave code \textsc{dacapo} using the RPBE exchange-correlation (xc)-functional \cite{RPBE,ASE}, converged plane wave cutoffs of 350 eV and 400 eV for rutile and perovskite metal oxides respectively, along with a density cutoff of 500 eV. The occupation of the Kohn-Sham orbitals was calculated at \(k_B T \approx 0.1\) eV, with all energies extrapolated to $T = 0$ K. The product of the supercell dimensions and the 
number of \textbf{k}-points \cite{Monkhorst} 
is \(\gtrsim\) 25 \AA\ in all repeated directions. For the anti-symmetric perovskite surface slab calculations a dipole correction has been employed \cite{DipoleCorrection}.

We have performed structural relaxations until a maximum force below 0.05 eV/\AA\ was obtained.  At the same time we have minimized the strain on the unit cell in all periodically repeated directions, and employed more than 10 \AA\ of vacuum between repeated nanorods, nanotubes, and surface slabs.  Schematics of the TiO$_2$ (2,2) nanorod, (3,3) nanorod, and (110) surface, along with the SrTiO$_3$ (100) TiO$_2$ and SrO terminated surfaces \cite{termination} are shown in Fig.~\ref{Structures}. The supercells used have been repeated four times along the nanorod and nanotube axis in Fig.~\ref{Structures}(a) and \ref{Structures}(b), and four times in the surface plane for the surfaces shown in Fig.~\ref{Structures}(c) and \ref{Structures}(d).  For the TiO$_2$ (110) surface our four layer thick slab yields surface formation energies of 0.44 J/m$^2$ in reasonable agreement with the GGA \cite{PWXC} 12 layer thick slab value of 0.50 J/m$^2$ \cite{TiO2SurfaceEnergies}, and the B3LYP \cite{B3LYP} 9 layer thick slab value of 0.67 J/m$^2$ \cite{SnO2SurfaceEnergies}.  Differences amongst these values are attributable to the choice of xc-functional and number of layers, especially for uneven slab calculations, which tend to yield higher energies and converge slower \cite{TiO2SurfaceEnergies}.  As expected, these values are below the experimental nanoparticle surface energy of $\text{2.2}\pm\text{0.2}$ J/m$^2$ for TiO$_2$, as shown in Fig.~\ref{Heats}~\cite{SurfaceEnthalpiesTiO2}.

Spin polarized calculations have been performed only for CrO$_2$ and MnO$_2$ rutile metal oxides, while all perovskites have been calculated spin polarized.  However, spin only proved to be important for the energetics of CrO$_2$, MnO$_2$, SrCrO$_3$, SrMnO$_3$, SrFeO$_3$, SrCoO$_3$, and SrNiO$_3$.  A table of these results is provided in Ref.~\onlinecite{SupplementaryMaterial}.

Serious questions have been raised as to whether standard DFT calculatons can describe metal oxides with sufficient accuracy. For the purposes of the present study, it is worth noting that it has recently been demonstrated that DFT calculations do reproduce semi-quantitatively the formation energies of bulk rutile and perovskite metal oxides \cite{Martinez,Calle}.

The formation energies of the various surfaces and nanostructures from Fig.~\ref{Structures} are shown in Fig.~\ref{Heats}.  In the following we shall analyze the nature of the formation energies, and relate them to other characteristics of the materials.
An obvious choice of descriptor for the (110) metal oxide surface formation energy \(\Delta E_f^{(110)}\) might be the bulk heat of formation \(\Delta G_f^{\text{Bulk}}\), which is well described by DFT for rutile metal oxides \cite{Martinez}.  However, as shown in Fig.~\ref{Heats}(a), these quantities appear to be poorly correlated. We instead observe a strong correlation between \(\Delta E_f^{(110)}\) and the $d$-band center \(\varepsilon_d\) for the bulk rutile metal oxides, as shown in Fig.~\ref{Heats}(b).  
Here, \(\varepsilon_d\) is the average energy, relative to the Fermi level \(\varepsilon_F\), of the density of states (DOS) projected onto the metal's atomic $d$-orbitals in the metal oxide \(n_d(\varepsilon)\), so that \(\varepsilon_d \equiv \int (\varepsilon-\varepsilon_F) n_d(\varepsilon)d\varepsilon\) \cite{dband2007}.
For the nanorod and nanotube formation energies, we also find little correlation with \(\Delta G_f^{\text{Bulk}}\), but a strong correlation with \(\varepsilon_d\).  

This suggests metal oxide surface and nanostructure formation energies may be considered perturbations of the bulk metal oxide's electronic structure due to bond breaking.  In this case \(\Delta E_f\) should be well described by the fraction of metal--oxygen (M--O) bonds which are broken \(\Delta \chi\), times the M--O bond energy in the bulk \(\varepsilon_{\text{M--O}}\).  For the case of (110) surface formation \(\Delta \chi^{(110)} \approx \nicefrac{1}{6}\) per MO$_{\text 2}$, while for nanorod and nanotube formation \(\Delta \chi^{\text{NT}} \approx \nicefrac{2}{6} \approx \nicefrac{1}{3}\) per MO$_{\text 2}$.  Assuming \(\varepsilon_{\text{M--O}}\sim -\varepsilon_d\), we indeed find that \(\Delta E_f \approx -\Delta \chi \varepsilon_d\), as shown in Fig.~\ref{Heats}(c).

This correlation may be explained qualitatively by recalling that the $d$-band is most stable when the average energy of the DOS is \(\varepsilon_F\), i.e.~\(\varepsilon_d \approx 0\).  In most cases this is equivalent to having a half full $d$-band \cite{dband2007}.  This is nearly the case for TiO$_2$, where \(\varepsilon_d \approx -0.67\) eV, and the (110) surface becomes very stable (\(\Delta E_f^{(110)} \approx 0.046\) eV/TiO$_2$).  As the $d$-band shifts down in energy, the energy stored in the M--O bonds increases accordingly, so that \(\varepsilon_{\text{M--O}} \sim -\varepsilon_d\).  

A more detailed analysis based on the molecular orbitals of the bulk metal oxide is presented in Refs.~\onlinecite{SupplementaryMaterial,ActaCrystalA,RutileBandStructure,GoodenoughReview}.
To summarize, as 
the number of $d$ electrons $N_d$ increases,
 the M--O coordinately unsaturated (cus) bond becomes stronger.  It is only this bond which is broken during (110) surface, nanorod and nanotube formation for rutile metal oxides.  As shown in 
Ref.~\onlinecite{SupplementaryMaterial}, 
this occurs as we progress from non-cus bonding ($N_d = $2),  M--M bond distortions ($N_d = $3), M--O cus bonding ($N_d = $4), non-bonding M orbitals with reduced distortions ($N_d = $5,6), to stronger M--O $p_\sigma$ bonds ($N_d = $7).  Further, for 
 a fixed number of $d$ electrons
we also find that the bond strength increases as the DOS shifts down in energy.  In effect, the energy stored in rutile metal oxide M--O bonds is then equal to the energy cost for ``shifting'' the bonding metal orbitals so that their DOS is symmetric about the Fermi level.  In such a case, both bonding with cus O atoms and distortions of the octahedral structure are minimized.

It is worth noting that we obtain near quantitative agreement between the DFT calculated formation energies and this simple model, when applied to naturally occurring transition metal rutile oxides.  We now show how this model \(\Delta E_f^{\text{Model}} \equiv -\Delta\chi\varepsilon_d\) may be extended to post transition metal oxides (GeO$_2$, SnO$_2$, and PbO$_2$) and 3$d$ transition metal perovskites (SrMO$_3$), as shown in Fig.~\ref{DFT_vs_model}.  

For post transition metals, the $d$-band is fully occupied, and does not participate in the M--O bonding.  Instead, the M--O bonding should occur via the occupied portion ($\nicefrac{1}{3}$) of the metal's $p$-band.  In any case, we find that the M--O bond energy \(\varepsilon_{\text{M--O}}\) for post transition metal oxides is approximately negative one third the $p$-band center \(-\varepsilon_p/3\), so that \(\Delta E_f \approx -\Delta\chi\varepsilon_p/3\). This approximation yields near-quantitative formation energies for post transition metal nanorods, nanotubes, and (110) surfaces, as seen in Fig.~\ref{DFT_vs_model}(a), (b), and (c), respectively.  

Since perovskite metal oxides share the MO$_6$ octahedral structure of rutile metal oxides, with interstitial Sr \cite{Perovskites,Cheng20051921}, it is reasonable to expect that the same correlation with the metal $d$-band centers \(\varepsilon_d\) should hold. From the DFT calculations for the perovskite (100) surface, we obtain the average of the formation energies for both the MO$_2$ and SrO terminated surfaces.  As such, our model predicts \(\Delta E_f^{(100)} \approx -(\Delta \chi^{\text{M}} \varepsilon_d^{\text{M}} + \Delta \chi^{\text{Sr}} \varepsilon_d^{\text{Sr}})/2\), where \(\Delta \chi^{\text{M}}\approx \nicefrac{2}{6} \approx \nicefrac{1}{3}\) per SrMO$_3$ and \(\Delta \chi^{\text{Sr}}\approx \nicefrac{2}{12}\approx\nicefrac{1}{6}\) per SrMO$_3$.
From Fig.~\ref{DFT_vs_model}(d), we find this is indeed the case.

In Fig.~\ref{DFT_vs_model}(a), (b), and (c) we also plot the DFT calculated formation energies versus \(\Delta E_f^{\text{Model}}\) for transition metals whose most stable metal oxide phase is \emph{not} rutile.  For these compounds, denoted by open symbols in Fig.~\ref{DFT_vs_model}, we find our model typically overestimates formation energies.  

For the HexABC layer structure of PtO$_2$ \cite{TiO2} we find the formation energies are completely independent of \(\varepsilon_d\) and \(\varepsilon_p\), as shown in 
Ref.~\onlinecite{SupplementaryMaterial}.  
However, this is also predicted by our model, since no M--O bonds are broken ($\Delta\chi \approx 0$) when forming the HexABC layer.  Instead, the energy cost to form the HexABC layer is related to the preference of oxygen to be either $sp^2$ or $sp^3$ hybridized in the metal oxide.

In conclusion, we have demonstrated that the formation energies for nanotubes, nanorods, and surfaces of metal oxides may be determined semi-quantitatively from the fraction of M--O bonds which are broken \(\Delta \chi\) and the bonding band centers \(\varepsilon_d\) and \(\varepsilon_p\) in the bulk metal oxide.  We anticipate such models will prove useful in predicting the formation energy of doped metal oxide surfaces and nanostructures, their reactant adsorption energies, and their activities. 

\acknowledgments

The authors acknowledge financial support from the Danish
Center for Scientific Computing and NABIIT. J.I.M. acknowledges the financial support of the STREP EU APOLLON-B Project, 
 and F.C.-V. acknowledges financial support from the Strategic Electrochemistry Research Center (SERC). The Center for Atomic-scale Materials Design (CAMD) is sponsored by the Lundbeck Foundation.


\end{document}